\documentclass[12pt]{iopart}
\usepackage{iopams}
\usepackage{graphicx}
\usepackage{setspace}
\usepackage{longtable}
\usepackage{pifont}
\usepackage[pagewise]{lineno}

\newcommand{\mmj}{minimax$|j|$}

\begin{document}
%\linenumbers
\title[Minimax Current Density Coil Design]{Minimax Current Density Coil Design}
\author{Michael Poole$^1$, Pierre Weiss$^2$, Hector Sanchez Lopez$^1$, Michael Ng$^3$ and Stuart Crozier$^1$}
\address{$^1$ School of Information Technology and Electrical
Engineering, University of Queensland, St. Lucia, Brisbane, QLD
4072, Australia.}
\address{$^2$ Institut de Mathematiques, Universit\'{e} Paul
Sabatier Toulouse 3, 31062, Toulouse, France.}
\address{$^3$ Department of Mathematics, Hong Kong Baptist
University, Kowloon Tong, Hong Kong.} \ead{michael@itee.uq.edu.au}
\begin{abstract}
``Coil design'' is an inverse problem in which arrangements of
wire are designed to generate a prescribed magnetic field when
energised with electric current. The design of gradient and shim
coils for magnetic resonance imaging (MRI) are important examples
of coil design. The magnetic fields that these coils generate are
usually required to be both strong and accurate. Other
electromagnetic properties of the coils, such as inductance, may
be considered in the design process which becomes an optimisation
problem. The maximum current density is additionally optimised in
this work and the resultant coils are investigated for performance
and practicality. Coils with minimax current density were found to
exhibit maximally spread wires and may help disperse localized
regions of Joule heating. They also produce the highest possible
magnetic field strength per unit current for any given surface and
wire size. Three different flavours of boundary element method
that employ different basis functions (triangular elements with
uniform current, cylindrical elements with sinusoidal current and
conic section elements with sinusoidal-uniform current) were used
with this approach to illustrate its generality.
\end{abstract} 
 \submitto{\JPD} \maketitle

\section{Introduction}\label{sec:intro}

In magnetic resonance imaging (MRI) an exquisitely uniform and
very intense magnetic field is used to polarize the spin
population of a sample so as to maximise the strength of the
nuclear magnetic resonance (NMR) signal. A process known as
``shimming'' is performed at the start of every scan to ensure
that this field is as uniform as possible. Shimming involves
adjusting the electric current in a set of ``shim coils'' that
each generate a magnetic field of spherical harmonic intensity in
the region of interest (ROI). MR images are formed by
superimposing magnetic field gradients which causes the
frequency of the NMR signal, the Larmor frequency, to vary linearly across
the sample. Fourier techniques reconstruct the image from these
frequency encoded NMR signals. The linearly varying magnetic
fields are generated by ``gradient coils''. This paper deals with
the design of both gradient and shim coils that dictate the speed,
resolution and accuracy of MRI \cite{Jin1998}.

Gradient and shim coil design is an inverse problem in which
arrangements of wire are required to generate a specified magnetic
field when energized. Additional considerations are required such
as minimal stored energy, so that they may be switched rapidly, or
minimal resistive power dissipation, so that their temperature
does not increase excessively. As MRI machines get shorter to
improve patient comfort \cite{Crozier1997a} so too must the
gradient coils \cite{Shvartsman2005}. Reducing the length of the
gradient coils pushes the wires closer together to maintain
magnetic field accuracy. However, gradient and shim coils are
constructed from finite sized wire and so there is a minimum wire
separation that can be built. In this work the maximum current
density was minimized in the coil design process which maximally
increases the minimum wire spacing of a coil for fixed coil
surface geometry. For a given engineering limit for the minimum
spacing between wires this technique can be used to increase the
efficiency of the coil (the amount of field per Amp\`{e}re). It
can also be used to reduce the \emph{local} power dissipation and
disperse the hot spots of a coil. The present study demonstrates
the design of coils with some increase in inductance or resistance
in order to spread wires. Such designs should be judged by
appropriate metrics that better encapsulate the coil design
problem than those designed to reflect purely the stored energy or
power dissipation of the design.

Coil design is also known as magnetic field synthesis and is
described by a Fredholm equation of the first kind, which is known
to be ill-posed \cite{Adamiak1981}. Early approaches to coil
design in MRI cancelled undesired spherical harmonic components of
the magnetic field by symmetry and appropriate positioning of
loops and arcs of wire (e.g. \cite{Romeo1984}) or by parameterized
surface current densities \cite{Golay1958}. The ``target field''
method \cite{Turner1986} circumvented the problem of ill-posedness
by employing a Fourier-Bessel expansion of the
$\frac{1}{|\mathbf{r}-\mathbf{r}^\prime|}$ Green's function (see
\S 3.11 in Ref. \cite{Jackson1998}) and defining a continuous
target field function for $-\infty < z < \infty$. Since Fourier
transforms have unique inverses, it is possible to analytically
determine the current density on an infinitely-long cylinder for a
limited set of target field functions.

If the target field is defined at a finite set of points, there
exists an infinite number of current densities that can produce
such a field. With the introduction of the minimum inductance as a
constraint \cite{Turner1988} the problem becomes regularized and a
unique solution exists. This is essentially Tikhonov
regularization \cite{Tikhonov1977} of the ill-conditioned system
of linear equations \cite{Forbes2001, Forbes2002, Forbes2003}.
This minimum inductance solution was found to be somewhat
impractical, so the field was allowed to deviate from its
prescribed target values in order to permit a smoother wire
pattern. In a related method, a smooth current density was always
obtained if it was defined as a weighted sum of a finite number
of truncated sinusoidal functions \cite{Stekly1985, Carlson1992,
Forbes2001, Forbes2002, Forbes2003}. Parameterizing the current
density in this manner allowed more practical, finite-length
cylindrical coils with limited spatial frequency to be designed in
a method sometimes referred to as the Turner-Carlson method.
Minimization of the inductance (the current-normalized stored
energy) can be easily substituted by the resistance
(the current-normalized power dissipation) while still resulting in a
unique solution \cite{Turner1993}. For
coil designs with asymmetric target field location it is necessary
to enforce zero net torque in the presence of an intense
background magnetic field \cite{Abduljalil1994}.

A coil may be defined by its surface current density. The
magnitude of that current density defines the wire spacing and the
amount of heat generated at positions on the surface. To spread
the closest wires or to reduce the local heating, the maximum
value of the current density magnitude was incorporated into the
coil design problem and its maximum value was minimised. This term
is not linear nor quadratic, but only convex with respect to the
current density. In contrast to more traditional approaches where
only linear systems are solved, we used techniques of convex
programming to handle the non-linearities and singularities that
arise from the ``max'' term. We developed an original algorithm to
solve this optimisation problem that can be seen as a continuation
of two works by Y. Nesterov \cite{Nesterov2005, Nesterov2007}. It
can be shown to converge to the global minimizer of the cost
function but details of this algorithm will be presented
elsewhere.

The concept of minimum maximum current density (\mmj) coil design
is general and is not limited to any particular coil design
method. In this work, three different boundary element methods
(BEM) were used to investigate the behaviour of \mmj\ coils. These
are the Turner-Carlson \cite{Stekly1985, Carlson1992}, triangular
\cite{Pissanetzky1992} and axisymmetric \cite{Peeren2003a,
Poole2008a} BEMs. The target field was specified at a finite set
of discrete points in a region of interest (ROI). The field
synthesis problem is defined as minimizing the sum-of-squares
field error and is an ill-posed problem. Therefore, a regularizing
term must be included to obtain a unique solution.

In a previous attempt to reduce the maximum current density the
regularization term was adaptively modified \cite{Poole2008b}.
This method showed a considerable reduction in maximum current
density, but it was not known how optimal the solutions were.
Other approaches in which regions of the coil were designed
manually have been used to control the maximum current density:
for example, by predefining the return conductors
\cite{Compton1984, Konzbul1995} or by manually introducing a large
number of constraints (page 146 of Ref.
\cite{Peeren2003a}). The method presented here truly
minimizes the maximum current density for coils designed on
surfaces of arbitrary shape that generate any physically
realizable magnetic field.

\section{Methods}\label{sec:methods}

\subsection{Physical Model}\label{sec:problem}

In magnetostatics, Amp\`{e}re's Law, $\nabla \times
\mathbf{B}(\mathbf{r}) = \mu_0 \mathbf{J}(\mathbf{r})$, relates
the magnetic field, $\mathbf{B}(\mathbf{r})$, and the free current
density, $\mathbf{J}(\mathbf{r})$. Current density must be
conserved, so $\nabla \cdot \mathbf{J}(\mathbf{r}) = 0$. Employing
the magnetic vector potential, $\mathbf{A}(\mathbf{r}) $, where
$\mathbf{B}(\mathbf{r}) = \nabla \times \mathbf{A}(\mathbf{r})$,
Amp\`{e}re's Law becomes a Poisson equation, $\nabla^2
\mathbf{A}(\mathbf{r}) = \mu_0 \mathbf{J}(\mathbf{r})$, which has
the solution

\begin{equation}\label{eq:A}
    \mathbf{A}(\mathbf{r}) =
    \frac{\mu_0}{4\pi}\int\!\!\frac{\mathbf{J}(\mathbf{r}^\prime)}
    {| \mathbf{r}-\mathbf{r}^\prime|}dV^\prime
\end{equation}

\noindent in the Coulomb gauge ($\nabla \cdot
\mathbf{A}(\mathbf{r}) = 0$). $\mu_0$ is the permeability of
free-space and has the value $4 \pi \times 10^{-7}$
Hm$^{-1}$. With some algebra, this leads to the familiar
volumetric integral form of the Biot-Savart law
\cite{Jackson1998},

\begin{equation}\label{eq:BS}
    \mathbf{B}(\mathbf{r}) =
    \frac{\mu_0}{4\pi}\int\mathbf{J}(\mathbf{r}^\prime) \times
    \frac{( \mathbf{r}-\mathbf{r}^\prime)}
    {| \mathbf{r}-\mathbf{r}^\prime|^3} dV^\prime.
\end{equation}

For each directional component of $\mathbf{B}(\mathbf{r})$,
\eref{eq:BS} is a Fredholm equation of the first kind which is
known to be ill-posed. MRI conventionally requires a strong
magnetic field for polarization of the nuclear spin states within
the sample to be imaged \cite{Jin1998}. We consider a system
immersed in a background magnetic field,
$\mathbf{B}_0(\mathbf{r})$, that is highly uniform, unidirectional
and very strong, i.e. $\mathbf{B}_0(\mathbf{r}) = B_{0z}
\hat{\mathbf{z}}$, where $\hat{\mathbf{z}}$ is the unit vector
parallel to the $z$-axis. The magnitude of the combined magnetic
field, $|B_{0z} \hat{\mathbf{z}} + \mathbf{B}(\mathbf{r})| \approx
(B_{0z} + B_z(\mathbf{r}))$, dictates the local Larmor frequency
of the NMR signal and subsequent spatial localization. Therefore
we only need to design coils to generate a specific
$B_z(\mathbf{r})$, justifiably neglecting the two other
components, $B_x(\mathbf{r})$ and $B_y(\mathbf{r})$.

In the context of this paper, ``coil design'' is the inversion of
\eref{eq:BS} to design an arrangement of wires that, when
energized, form a current density which generates a
prescribed magnetic field. The region of space in which the field
is prescribed, the ROI, is separate from the region in which the
current density exists.

Coil design is rarely as simple as inverting \eref{eq:BS} but
requires the consideration of other electromagnetic properties.
The stored energy, $W$, associated with $\mathbf{J}(\mathbf{r})$
is \cite{Jackson1998}

\begin{equation}\label{eq:W}
W = \frac{\mu_0}{8\pi}\int_{\Omega_c}\!\!\int_{\Omega_c}
\frac{\mathbf{J}(\mathbf{r})\cdot\mathbf{J}(\mathbf{r}^\prime)} {|
\mathbf{r}-\mathbf{r}^\prime|}dV dV^\prime,
\end{equation}

\noindent where $\Omega_c$ is the region of the coil in which
$\mathbf{J}(\mathbf{r})$ is confined to flow. The resistive power
dissipation, $P$, is

\begin{equation}\label{eq:P}
    P = \rho_{Cu} \int_{\Omega_c} |\mathbf{J}(\mathbf{r})|^2 dV,
\end{equation}

\noindent where $\rho_{Cu}$ is the resistivity of the conducting
medium which, in this case, is assumed to be copper, $\rho_{Cu} =
1.68 \times 10^{-10} \Omega$ m.

The coil may be in close proximity to other conducting surfaces
defined by the region $\Omega_e$. Changing
$\mathbf{J}(\mathbf{r})$ in time causes $\mathbf{B}(\mathbf{r})$
to also change, Faraday's Law, $\nabla \times
\mathbf{E}(\mathbf{r}) = \frac{\partial
\mathbf{B}(\mathbf{r})}{\partial t}$, and $\mathbf{J}(\mathbf{r})
= \sigma \mathbf{E}(\mathbf{r})$ show that currents may be induced
in other conducting surfaces. These ``eddy currents'' can cause
deleterious effects on MRI. So, a coil designer must consider the
effects that the induced eddy currents have on the field in the
ROI \cite{Mansfield1986a}. For low frequencies ($<$10 kHz) the
quasistatic approximation may be used and following the approach
of Peeren \cite{Peeren2003}, a Heaviside function response in the
coil current was assumed. This leads to a linear relationship
between coil currents and eddy currents.

Lorentz forces act on the coil when immersed in a background
magnetic field, $\mathbf{B}_0$. The net Lorentz force is zero for
a divergence-free current density in a uniform $\mathbf{B}_0$, but
there may exist a consequential net torque, $\btau$,

\begin{equation}\label{eq:M}
\btau = \int_{\Omega_c} \mathbf{r} \times \left[
\mathbf{J}(\mathbf{r}) \times \mathbf{B}_0\right] dV.
\end{equation}

Current density was confined to flow on thin surfaces so that a
scalar stream-function, $\psi(\mathbf{r})$, can be used to define
the vector current density

\begin{equation}\label{eq:sfdef}
\mathbf{J}(\mathbf{r}) = \nabla \times \left[\psi(\mathbf{r})
\hat{\mathbf{n}}(\mathbf{r})\right],
\end{equation}

\noindent where $\hat{\mathbf{n}}(\mathbf{r})$ in the unit vector
normal to the surface at $\mathbf{r}$.

\subsection{Discrete Formulation}\label{sec:disc}

The coil design problem may be solved analytically for some
special cases \cite{Turner1986} but for other geometries the
physical problem must be described by a finite number of
parameters in order to apply numerical methods. The type of
parameterisation may chosen to best suit the type of coil that is
to be designed. $\psi(\mathbf{r})$ can be approximated as a finite
weighted sum of $N$ basis functions,

\begin{equation}\label{eq:sfdisc}
    \psi(\mathbf{r}) \approx \sum_n^N \psi_n
    \hat{\psi}_n(\mathbf{r}),
\end{equation}

\noindent and so can the current density by combination of
\eref{eq:sfdisc} with \eref{eq:sfdef},

\begin{equation}\label{eq:jdisc}
    \mathbf{J}(\mathbf{r}) \approx \sum_n^N \psi_n
    \hat{\mathbf{j}}_n(\mathbf{r}),
\end{equation}

\noindent where $\hat{\psi}_n(\mathbf{r})$ and
$\hat{\mathbf{j}}_n(\mathbf{r})$ are the $n$th stream-function and
current density basis functions respectively, and $\psi_n$ are the
weights.

\Eref{eq:jdisc} can be incorporated in \eref{eq:BS} to
\eref{eq:M}, so that $B_z(\mathbf{r})$, $W$, $P$ and $\btau$ are
parameterised as finite summations.

\subsection{Matrix Equations}

This discrete formulation allows matrix equations for each of the
physical properties to be written. A vector of stream-function
weights, $\psi$, was defined; $\psi = \left[
\psi_1,\ldots,\psi_n,\ldots,\psi_N \right]^T$ (where $^T$
represents the transpose operation). Each Cartesian component of
the current density at a set of points, $\mathbf{r}_s$, can be
written as matrix equations

\begin{equation}\label{eq:matG}
    j_x = J_x \psi, \quad j_y = J_y \psi, \quad  j_z = J_z \psi,
\end{equation}

\noindent where $j_x$ is a vector that lists values of the
$x$-component of the current density at a set of $S$ points, $j_x
= \left[
J_{x}(\mathbf{r}_1),\ldots,J_{x}(\mathbf{r}_s),\ldots,J_{x}(\mathbf{r}_S)
\right]^T$ and $J_x$ in an $S \times N$ matrix. Similar matrix
equations can be written for the cylindrical coordinate system to
give $j_\rho$, $j_\phi$ and $j_z$.

A $H \times N$ matrix $B$ relates $\psi$ to a vector $b$ of length
$H$ containing magnetic field values, where $b = \left[
B_{z}(\mathbf{r}_1), \ldots, B_{z}(\mathbf{r}_h), \ldots,
B_{z}(\mathbf{r}_H) \right]^T$;

\begin{equation}\label{eq:matB}
    b = B \psi.
\end{equation}

Similarly, each component of the torque vector \eref{eq:M} can be
written as the inner product of $\psi$ and a vector,

\begin{equation}\label{eq:matM}
    \tau_x = T_x \psi, \quad\tau_y = T_y \psi, \quad\tau_z = T_z
    \psi.
\end{equation}

The energy terms \eref{eq:W} and \eref{eq:P} are quadratic with
respect to $\psi$,

\begin{equation}\label{eq:matL}
    W = \psi^T {L_c} \psi
\end{equation}

\begin{equation}\label{eq:matR}
    P = \psi^T {R_c} \psi
\end{equation}

\noindent where $L_c$ and $R_c$ are symmetric, $N \times N$
matrices of the inductance and resistance of the coil surface
respectively. In fact, $R_c = J_x^TJ_x + J_y^TJ_y + J_z^T J_z$.

Three types of parameterisation are used in this work. The first
assumes that the current-carrying surface is a finite-length
cylinder and that $\psi(\mathbf{r})$ is a weighted sum of
truncated sinusoidal functions \cite{Stekly1985,Carlson1992}. In
the second approach, surfaces are described by flat triangular
elements and $\psi(\mathbf{r})$ is a piecewise-linear function
\cite{Pissanetzky1992, Peeren2003, Lemdiasov2005, Poole2007}. The
third approach uses surfaces of revolution about the $z$-axis and
is an axisymmetric BEM \cite{Peeren2003a,Poole2008a}. The way in
which $\psi(\mathbf{r})$ and $\mathbf{J}(\mathbf{r})$ are
parameterised in each case are given in the Appendices. For
details of how to calculate the matrices, $B$, $T_x$, $T_y$,
$T_z$, $L_c$, $R_c$ for each type of parameterisation, the reader
is advised to seek the above references. Other approaches to
discretizing the problem are possible, such as using quadrilateral
elements, but are not described in the present work.

\subsection{Numerical Problem}

The vector $\psi$ is a list of the stream-function weights,
$\psi_n$, which are the free parameters of the coil design
problem. The problem including the maximum current density and all
other terms can be written generally

\begin{equation}\label{eq:problem}
\min_{\psi \in \Psi} \{ U(\psi) = f(\psi) + \alpha e(\psi) + \beta
W(\psi) + \gamma P(\psi) + \delta \|j(\psi)\|_\infty \}.
\end{equation}

\noindent It contains terms to control the residual primary field,
$f(\psi)$, eddy current field, $e(\psi)$, stored magnetic energy,
$W(\psi)$, power dissipation, $P(\psi)$, and maximum current
density, $\|j(\psi)\|_\infty$, along with their respective,
user-definable weighting factors, $\alpha$, $\beta$, $\gamma$ and
$\delta$. One, two or three of these parameters are usually set
equal to zero to remove them from $U(\psi)$. For example, $\gamma
= \delta = 0$ will result in an actively shielded, torque-balanced
coil with minimal stored energy. Each term in \eref{eq:problem}
possesses a natural scaling from the physical constants used in
their calculation. Choice of $\alpha$, $\beta$, $\gamma$ and
$\delta$ values must balance these scalings: for example,
$\alpha$, $\beta$, $\gamma$ and $\delta$ are typically in the
order 1, $10^{-7}$, $10^{-9}$ and $10^{-10}$ respectively so that
they have a magnitude comparable with the $f(\psi)$ term, but are
dependent on the specific problem.

The minimization was performed such that $\psi$, belonged to the
set of stream-functions, $\Psi$, that exhibit zero net torque
\eref{eq:matM};

\begin{equation}\label{eq:noTorqueSet}
    \Psi = \{ \psi \in \mathbb{R}^N, \quad T_x \psi = 0 \quad
    \& \quad T_y \psi = 0\},
\end{equation}

\noindent where $T_x \psi$ and $T_y \psi$ give the $x$- and
$y$-components of the torque vector $\tau_x$ and $\tau_y$,
respectively. In this work it was assumed that the background
magnetic field was uniform and oriented parallel to $z$, and as
such $\tau_z = 0$. It is also possible to balance the torque of
coils immersed in non-uniform background magnetic fields.

The $f$ term in \eref{eq:problem} represents the sum-of-squares of
the error in the primary magnetic field,

\begin{equation*}\label{eq:fieldErr}
    f = \frac{1}{2} \| B_c \psi - b_t\|_2^2,
\end{equation*}

\noindent where $\|\cdot\|_2$ is the classical $\ell^2$-norm,
$B_c$ is a $H \times N$ matrix relating $\psi$ to the magnetic
field values at the $H$ target field points in the ROI
\eref{eq:matB} and $b_t$ is a vector of length $H$ containing the
target magnetic field values.

The term $e$ in \eref{eq:problem} represents the sum-of-squares of
the magnetic field that the eddy currents produce in the ROI. A
Heaviside function in coil current was assumed \cite{Peeren2003}
and the stream-function of the instantaneously-induced eddy
current density, $\psi_e$, is linearly related to $\psi$ by

\begin{equation}\label{eq:psiEddy}
    \psi_e = -L_{e}^{-1} M_{ec} \psi,
\end{equation}

\noindent where $L_{e}$ is an $N_e \times N_e$ ($N_e$ is the
number of basis functions approximating the current density on the
eddy current surface) self-inductance matrix of the conducting
surface where eddy currents are induced and $M_{ec}$ is an $N_e
\times N$ matrix of the mutual inductance between the coil surface
and eddy current surface.

The field produced by the eddy current at the target points was
desired to be minimal, hence we use the sum-of-squares eddy
current field to enforce active magnetic shielding.

\begin{equation}\label{eq:eddyField}
    e = \frac{1}{2} \| B_e L_{e}^{-1} M_{ec} \psi\|_2^2,
\end{equation}

\noindent where $B_e$ is a $H \times N_e$ matrix relating $\psi_e$
to the eddy current magnetic field values at the target points.

The stored magnetic energy, $W$, and power dissipation, $P$, terms
in \eref{eq:problem} are quadratic with respect to $\psi$ and are
given by Equations \eref{eq:matL} and \eref{eq:matR},
respectively.

The maximum current density magnitude in the coil design is
written here as the $\ell^\infty$-norm of the current density
magnitude vector, $j$. $j$ is a list of length $S$ containing
the current density magnitude values at each surface point

\begin{equation}\label{eq:infNorm}
    \|j\|_\infty = \lim_{p\rightarrow \infty } \left( \sum_s (j_s)^p
    \right)^{1/p} := \max_s\{ j_s \},
\end{equation}

\begin{equation}\label{eq:currentDensityMagnitude}
    j_s = |\mathbf{j}_s| = \sqrt{j_{sx}^2 + j_{sy}^2 + j_{sz}^2}.
\end{equation}

\subsection{Optimization Algorithm}\label{sec:optimAlgo}

Previous methods solved $\min \{ U(\psi) \}$ by partial
differentiation of $U(\psi)$, $\frac{\partial U}{\partial \psi}$,
and subsequent matrix inversion of the consequential system of
linear equations \cite{Lemdiasov2005, Poole2007}. This cannot be
done with \eref{eq:problem} since $U(\psi)$ contains a
non-differentiable $\ell^\infty$ term. To solve \eref{eq:problem}
we used an accelerated descent algorithm of Nesterov
\cite{Nesterov2007} on a dual problem smoothed using ideas of
Moreau-Yosida. Full details of the algorithm will be submitted
elsewhere. For the purposes of this paper it should be noted that
the algorithm requires as inputs the smoothing parameter, $\mu$,
for the $\ell^\infty$-norm, the number of iterations to perform,
$Q$, and an initial guess for the solution, $\psi_0$. $\mu$ and
$Q$ are related by some inverse relationship that requires more
iterations when less smoothing is applied, but will approximate
the non-differentiable $\ell^\infty$-norm to a greater degree.
Convergence was checked by observing the value of the dual cost
function as $q \rightarrow Q$.

The optimization algorithm was coded in Matlab (The Mathworks,
Natick, MA) and was excecuted on a 64-bit Linux server with Intel
(Intel Corporation, Santa Clara, CA) Xeon E5430 quad-core CPUs at
2.66GHz.

\subsection{Examples}\label{sec:Examples}

The impact of designing coils with the \mmj\ was investigated by three
examples. These examples are described in the following sections
and were chosen to elucidate the behavior of the system when
designing realistic coils. In the examples outlined below the
convergence rates and calculation times were recorded.

Relevant properties of the coil performance were recorded in all
cases. The efficiency, $\eta$, is the intensity of magnetic field
that the coil can generate with 1 Amp and is also sometimes
referred to as the sensitivity. Inductance, $L$, resistance, $R$,
minimum spacing between wires, $w$, maximum field error in the
ROI, max($\Delta B_z$) and maximum eddy current field in the ROI,
max($B_{ez}$), are all recorded. Derived figures-of-merit (FOMs)
$\eta^2/L$, $\eta^2/R$ and $\eta w$ are independent of the number
of contours, $N_c$, used to convert $\psi$ into wires and are
useful for comparing between coils.

\subsubsection{Cylindrical X-gradient
coils}\label{sec:sosCylMeth}

We initially demonstrate \mmj\ coil design with $\psi(\mathbf{r})$
parameterised by a sum of sinusoidal basis functions
\cite{Stekly1985,Carlson1992}. \ref{sec:appA} details the
parameterisation of $\psi(\mathbf{r})$ and
$\mathbf{J}(\mathbf{r})$. The current-carrying surface for these
examples was assumed to be a finite length cylinder 760 mm in
diameter. The region of uniformity (ROU) was a 400 mm long, 400 mm
diameter cylindrical region positioned concentrically inside the
current-carrying surface. The target field in the ROI has a
magnitude that varies linearly in the $x$-direction;
$B_z(\mathbf{r}) \propto x$. This simple geometry was used to
investigate some of the more fundamental behaviours of coils
designed with \mmj. In all cases $\max{(\Delta B_z)}$ was kept at
$5 \pm 0.01$ \% and no torque balancing was required because
$\mathbf{J}(\mathbf{r})$ is forced to be symmetric by the limited
parameterisation. No active shielding was used, so $\alpha = 0$.

The length of the coil, $l$, was varied between 700 and 2000 mm to
observe its behavior with respect to standard min($W$) and
min($P$) coils.

In a second experiment, \mmj\ coils were designed with varying
amounts of power minimization to investigate their behavior on the
continuum from min($P$) to minimax$|j|$. This was performed with
$l$ = 1400 mm, for $N=36$ and 200. For $N=36$, with reference to
\eref{eq:sos0}, $M^\prime = 4$ and $N^\prime = 8$. For $N=200$,
$M^\prime = 10$ and $N^\prime = 20$.

\subsubsection{Shielded Gradient
Coils}\label{sec:methWBgrad}

Short, cylindrical, actively-shielded gradient coils were designed
with \mmj. The dimensions of a coil presented in Reference
\cite{Shvartsman2007} was used in this example. The system was
modelled with a triangular BEM \cite{Pissanetzky1992} which
approximates $\psi(\mathbf{r})$ as piecewise-linear in each
triangle as described in \ref{sec:appB}.

Four different X-gradient coils were designed using different
types of minimization, all with max$(\Delta B_z) = 5 \pm 0.01 \%$
in the ROI and $\alpha$ = 20; min($W$), min($P$), minimax$|j|$ and
min($P$ \& max$|j|$) which is some combination of min($P$) and \mmj.

A full set of gradient coils comprises X, Y and Z coils, with Y
being a $90^\circ$ rotation of the X-gradient. Z-gradient coils
($B_z(\mathbf{r}) \propto z$) were designed with min($P$), \mmj
and min($P$ and max$|j|$). It is known that for axi-symmetric
geometries and target fields (i.e. zonal coils) $\psi(\mathbf{r})$
is $\phi$ invariant. All nodes with identical $\rho_n$ and $z_n$
were treated together and forced to have the same value of
$\psi_n$. In some way this is equivalent to the axisymmetric case
described in the next example.

\subsubsection{Shim Coils}\label{sec:methShim}

Designing shim coils not only requires the production of magnetic
fields that have a different spatial form to gradient coils, but
the engineering and electronic requirements are also different.
The efficiency, $\eta$, is of primary importance and higher order
shims are considerably less efficient than low order shims.
Improving $\eta$ for the higher order shim coils would be useful
to improve correction of geometric distortion in MR images induced
by $\mathbf{B}_0$ field error and may provide smaller linewidths
for MR spectroscopy via higher order shimming. Due to the often
constrained axial and radial space provided for shim coils,
wire-spacing can become a problem and limits $\eta$. Coils
designed with the \mmj\ were studied to see if they could help
improve shim coil performance. X2-Y2 biplanar shim coils
($B_z(\mathbf{r}) \propto x^2 - y^2$) were designed with 860 mm
diameter and 500 mm separation \cite{Zhu2008}. The ROI is a
spherical volume of 380 mm diameter in which max($\Delta B_z$) was
fixed at $10 \pm 0.01$ \%. In both cases, the $\phi$ dependence of
$\psi(\mathbf{r})$ was spectrally decomposed in terms of 7
sinudoids, i.e. $M^\prime = 7$, see \ref{sec:appC}.

\section{Results}\label{sec:results}

\subsection{General Observations}\label{sec:genObservations}

Calculation of the system matrices for both the sum-of-sinusoids
and the axisymmetric BEM took on the order of a few seconds. The
time required for the triangular BEM system matrix calculations
was reduced over previously reported times \cite{Poole2007} by
coding this part in C. It took less than 10 minutes to calculate
all the matrices for a large problem containing 3200 nodes and
6144 triangles. For a medium-sized problem containing 1632 nodes
and 3072 triangles the system matrix calculation time was less
than 2 minutes.

The optimization algorithm converged in all cases as expected from
its deterministic nature. The smoothing parameter, $\mu$, for the
$\ell^\infty$-norm controls how close the solution to the smoothed
problem, $\psi_\mu^\ast$, is to the true solution, $\psi^\ast$.
Practically, we make $\mu$ small enough so that no observable
difference in the solutions is seen for smaller $\mu$. The time
required to find a solution close to $\psi^\ast$ varies widely and
is dependent on the problem. The number
of iterations, $Q$, that the algorithm required to converge is
inversely related to $\mu$. For a typical problem tackled in this
work, $1 \times 10^{-16} \lesssim \mu \lesssim 1 \times 10^{-14}$
resulted in indistinguishable solutions which typically required
$10,000 \lesssim Q \lesssim 200,000$.

\subsection{Cylindrical X-gradient
coils}\label{sec:resCyl}

\begin{figure}[h]
    \begin{center}
        \includegraphics[width=0.45\textwidth]{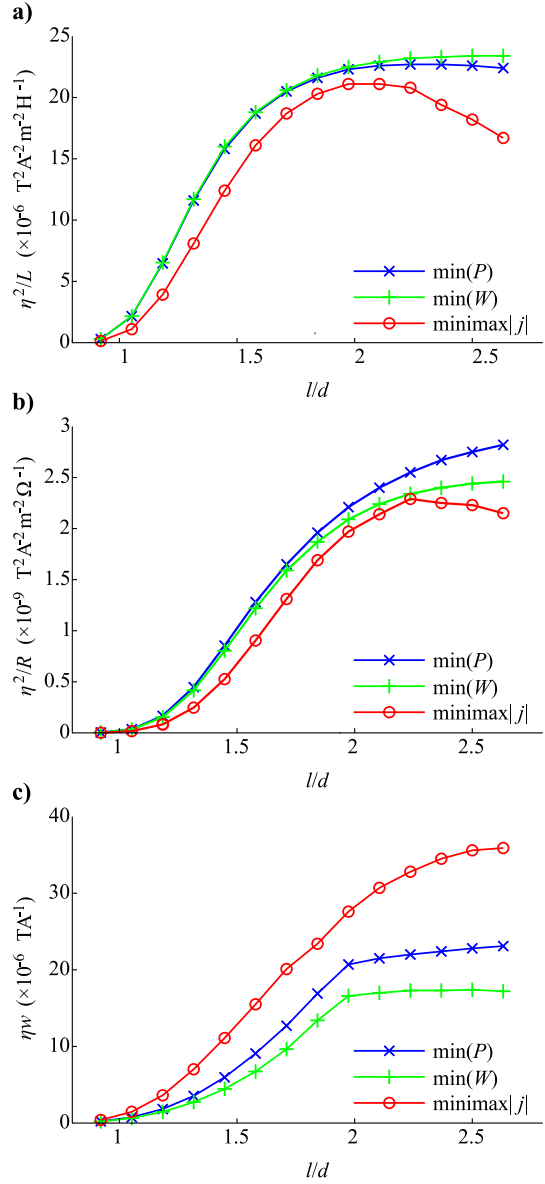}
    \end{center}
    \caption{ \label{fig:length} Variation of \textbf{a)} $\eta^2/L$, \textbf{b)} $\eta^2/R$ and
    \textbf{c)} $\eta w$ with length-to-diameter ratio, $l/d$ for
    unshielded X-gradient coils designed with a sum-of-sinusoids parametrised
    stream function.
    }
\end{figure}

\begin{figure}[h]
    \begin{center}
        \includegraphics[width=0.45\textwidth]{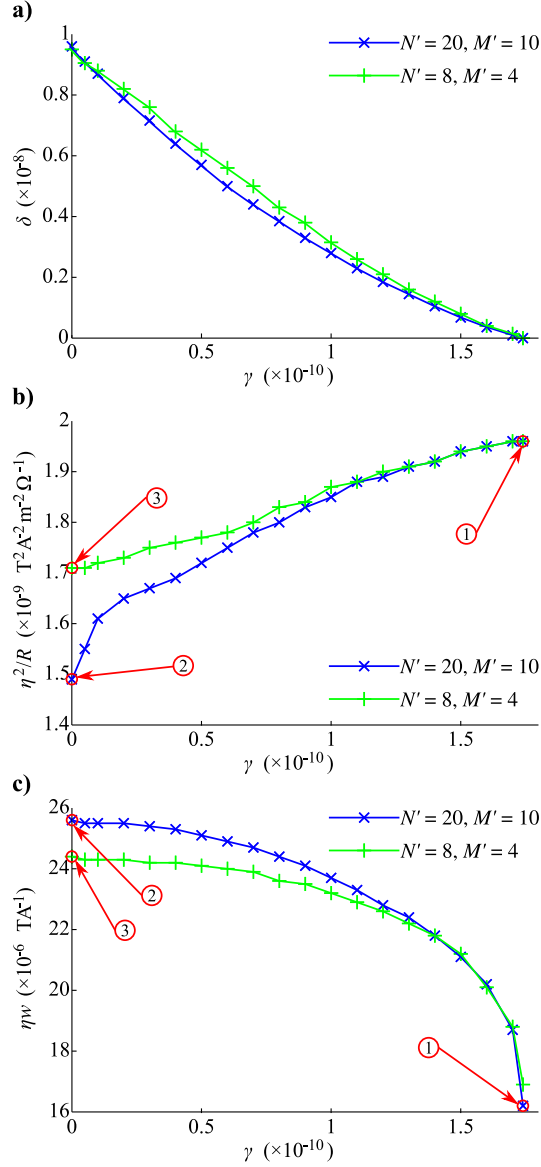}
    \end{center}
    \caption{ \label{fig:gammaDelta} \textbf{a)} $\delta$ values required
    in combination with varying $\gamma$ in order to maintain field
    error of $5 \pm 0.01$ \% with $\beta =0$. \textbf{b)} the resulting $\eta^2/R$ and \textbf{c)} $\eta w$
    of the coils for 32 and 200 sinusoidal basis functions. Data points labelled \ding{172}, \ding{173} and \ding{174} correspond to the coils shown in Figure \ref{fig:sosgrads} b), c) and d) respectively.
    }
\end{figure}

\begin{figure}[h]
    \begin{center}
        \includegraphics[width=\textwidth]{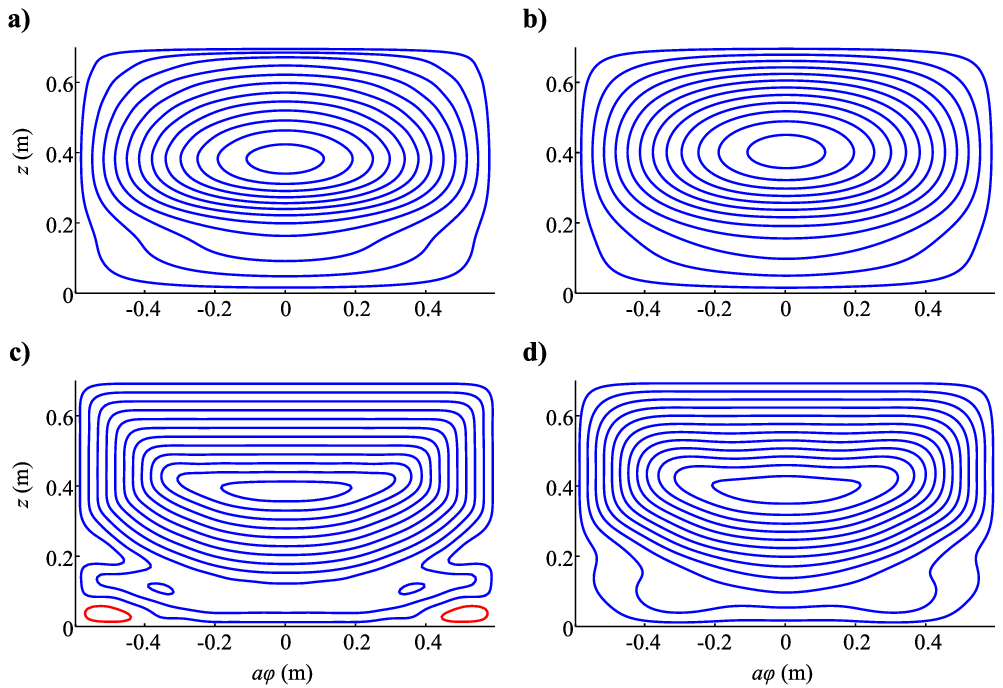}
    \end{center}
    \caption{ \label{fig:sosgrads} One quadrant each of the wire paths for the sum-of sinusoids, unshielded,
    X-gradient coils in the cases of \textbf{a)} min($W$), \textbf{b)} min($P$), \textbf{c)} \mmj\ with 200 sinusoids and \textbf{d)} \mmj\ with 32 sinusoids.
    Red wires indicate reversed current flow with respect to blue and only 12 contours of the stream function are shown for clarity.
    }
\end{figure}

The time required to find the solution to a small problem with
$\mu = 1 \times 10^{-15}$, $Q = 20,000$ and the number of free
variables, $N = 36$ was approximately 23 seconds. Figure
\ref{fig:length} shows how the FOMs for a) stored energy, b) power
dissipation and c) wire spacing varied for min($W$), min($P$) and
\mmj\ X-gradient coils as the length of the coil surface varied
with max$(\Delta B_z) = 5 \pm 0.01 \%$. Figure
\ref{fig:gammaDelta} a) shows the values of $\delta$ that were
used with varying $\gamma$ to maintain max$(\Delta B_z) = 5 \pm
0.01 \%$ in the ROU. The variation of the two relevant FOMs with
$\gamma$ are shown in Figure \ref{fig:gammaDelta} b) and c).
Figure \ref{fig:sosgrads} shows one quadrant of the wire paths for
four coils designed with a) min($W$), b) min($P$), c) \mmj\ with N
= 200 and d) \mmj\ with N=36. Wire positions are unwrapped from
their cylindrical shape onto a flat $z$-$a\phi$ plane. As with all
coils presented in this paper, connections must be made during
construction from each loop to its neighbour to ensure current
flow throughout the coil. The location of these coils are marked
for reference on Figure \ref{fig:gammaDelta} b) and c) where
\ding{172}, \ding{173} and \ding{174}  to the coils in Figure
\ref{fig:sosgrads} b), c) and d) respectively.

\subsection{Shielded Gradient
Coils}\label{sec:resultsWBgrad}

\begin{figure}[h]
    \begin{center}
        \includegraphics[width=\textwidth]{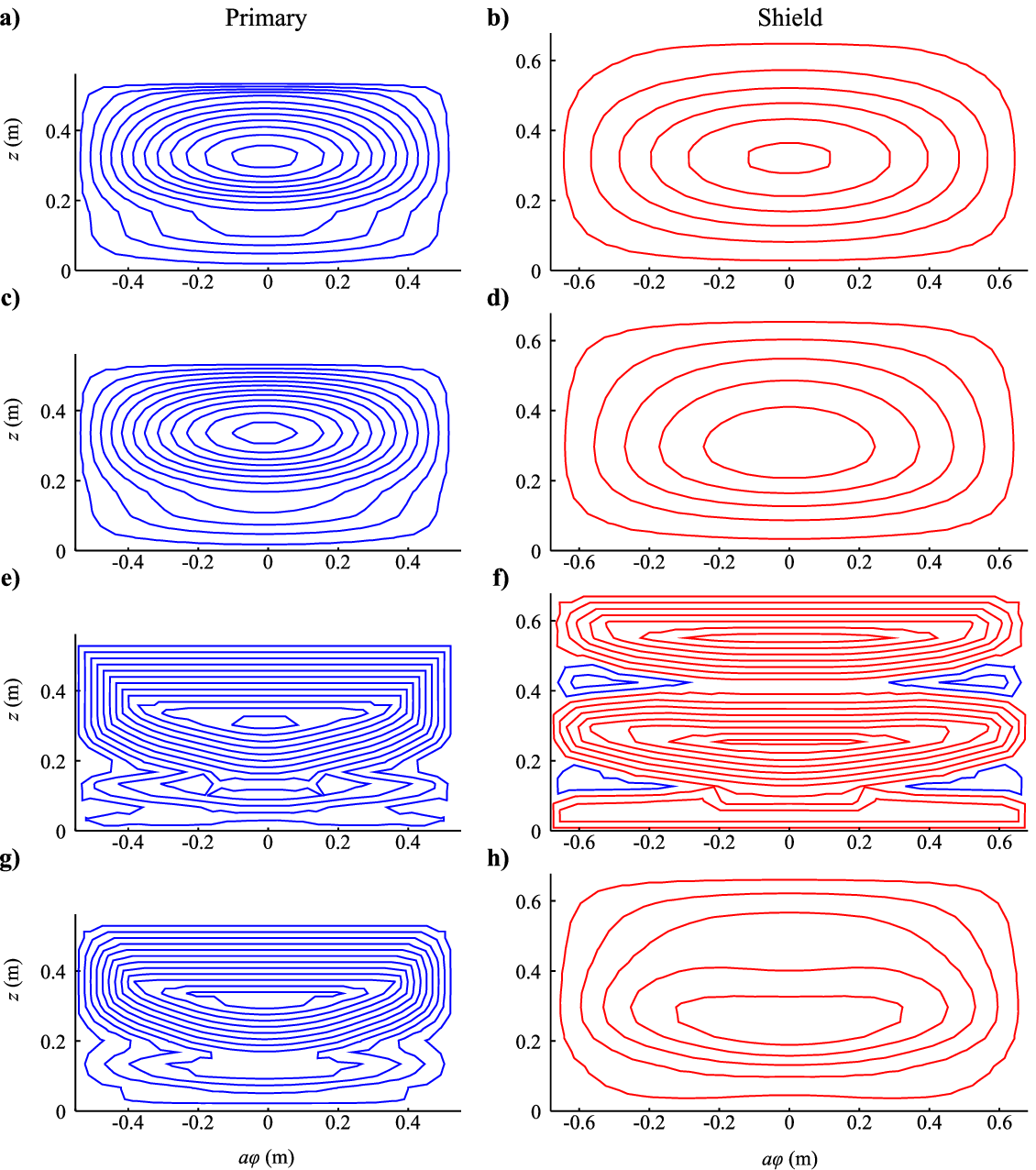}
    \end{center}
    \caption{ \label{fig:wbxgrad} Wire paths for the triangular
    BEM, actively-shielded, X-gradient coils in the cases of \textbf{a)} and \textbf{b)} min($W$), \textbf{c)} and \textbf{d)} min($P$), \textbf{e)} and \textbf{f)}
    \mmj\ and \textbf{g)} and \textbf{h)} min($P$ \& max$|j|$). The active magnetic screens appear on the right for the primary coils on the left.
    Red wires indicate reversed current flow with respect to blue and only 12 contours of the stream function are shown for clarity.
    }
\end{figure}

One quadrant of the wire paths for min($W$), min($P$), \mmj, and
min($P$ \& max$|j|$) shielded X-gradient coils are shown in Figure
\ref{fig:wbxgrad}. The left-hand side show the primary coils
and the right-hand side are their active magnetic shields. Their
performance characteristics are given in Table \ref{tab:results}.
Due to the high number of free-parameters, $N = 1985$, it took
approximately 480 minutes to perform $Q = 210,000$ iterations to
obtain a well converged solution.

\begin{table}
\caption{\label{tab:results} Properties of the triangular BEM
designed, actively-shielded X-gradient coils. Input parameters
$\alpha$, $\beta$, $\gamma$, $\delta$ and coil properties
including efficiency, $\eta$, maximum field error in the ROU,
max($\Delta B_z$), maximum eddy current field in the ROU,
max($B_{ez}$), inductance, $L$, resistance, $R$, minimum wire
spacing, $w$ and figure-of-merit values, $\eta^2/L$, $\eta^2/R$,
$\eta w$.} \footnotesize\rm
\begin{tabular*}{\textwidth}{@{}l*{15}{@{\extracolsep{0pt plus12pt}}l}}
    \br
Property & min($W$) & min($P$) & \mmj\ & min($P$ \& max$|j|$) \\
\mr
Coil & A & B & C & D\\
$\alpha$ & 20 &20&20&20\\
$\beta$ & 1.54$\times10^{-7}$ &0&0&0\\
$\gamma$ & 0 &5.5$\times10^{-9}$&0&$0.5 \times 10^{-9}$\\
$\delta$ & 0 &0&6.3$\times10^{-9}$&$5.1 \times 10^{-9}$\\
\mr
$\eta$, ($\mu$Tm$^{-1}$A$^{-1}$) & 72.9 & 71.7 & 74.0 & 74.3\\
max($\Delta B_z$) (\%) & 5.0 & 5.0 & 5.0 & 5.0\\
max($B_{ez}$) (\%) & 0.11 & 0.09 & 0.16 & 0.13 \\
$L$ ($\mu$H) & 661 & 654 & 1621 & 843 \\
$R$ (m$\Omega$) & 132 & 111 & 392 & 158 \\
$w$ (mm) & 4.1 & 6.0 & 9.3 & 9.0 \\
$\eta^2/L$ (T$^2$m$^{-2}$A$^{-2}$H$^{-1}$) & 8.3$\times10^{-6}$& 8.1$\times10^{-6}$& 5.4$\times10^{-6}$& 6.6$\times10^{-6}$\\
$\eta^2/R$ (T$^2$m$^{-2}$A$^{-2} \Omega ^{-1}$) & 4.0$\times10^{-8}$& 4.6$\times10^{-8}$& 1.4$\times10^{-8}$& 3.5$\times10^{-8}$\\
$\eta w$ (TA$^{-1}$) & 3.0$\times10^{-7}$& 4.3$\times10^{-7}$& 6.9$\times10^{-7}$& 6.7$\times10^{-7}$\\
\br
\end{tabular*}
\end{table}

\begin{figure}[h]
    \begin{center}
        \includegraphics[width=0.5\textwidth]{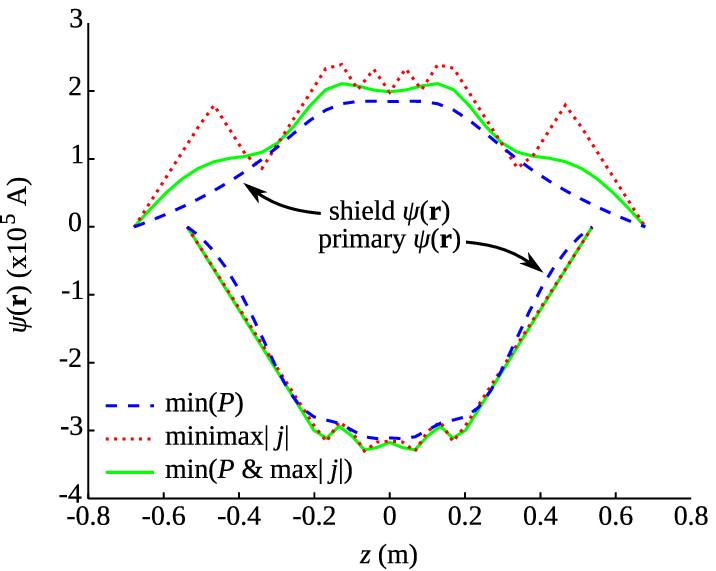}
    \end{center}
    \caption{ \label{fig:wbzgrad} Stream-functions for the
    actively-shielded, Z-gradient coils designed with
    min($P$), \mmj\ and min($P$ \& max$|j|$). The shield stream functions are greater than
    zero and the primary stream function are less than zero.
    }
\end{figure}

The stream-functions of the current densities of the three
Z-gradient coils with min($P$), \mmj, and min($P$ \& max$|j|$) are
shown in Fig. \ref{fig:wbzgrad}. The FOMs are given in Table
\ref{tab:Z}. The calculation time required to find the optimal
solution was dramatically reduced for the Z-gradient coils by
reducing the number of free variables. The time required for $Q =
80,000$ iterations was 10 minutes for a well converged solution.

\begin{table}
\caption{\label{tab:Z}Properties of the triangular BEM designed,
actively-shielded Z-gradient coils. Input parameters $\alpha$,
$\beta$, $\gamma$, $\delta$ and figure-of-merit values,
$\eta^2/L$, $\eta^2/R$, $\eta w$ are given.} \footnotesize\rm
\begin{tabular*}{\textwidth}{@{}l*{15}{@{\extracolsep{0pt plus12pt}}l}}
    \br
Property & min($P$) & \mmj\ & min($P$ \& max$|j|$) \\
\mr
$\alpha$ & 20 &20&20\\
$\beta$ & 0 &0&0\\
$\gamma$ & 6.1$\times10^{-9}$ &0&5$\times10^{-10}$\\
$\delta$ & 0 &7.8$\times10^{-9}$&7.6$\times10^{-9}$\\
\mr
$\eta^2/L$ (T$^2$m$^{-2}$A$^{-2}$H$^{-1}$)      & 1.35$\times10^{-5}$& 0.73$\times10^{-5}$& 1.16$\times10^{-5}$\\
$\eta^2/R$ (T$^2$m$^{-2}$A$^{-2} \Omega ^{-1}$) & 1.44$\times10^{-8}$& 0.60$\times10^{-8}$& 1.05$\times10^{-8}$\\
$\eta w$ (TA$^{-1}$)                            & 8.45$\times10^{-7}$& 11.43$\times10^{-7}$& 11.13$\times10^{-7}$\\
\br
\end{tabular*}
\end{table}

\subsection{Shim Coils}\label{sec:resShim}

\begin{figure}[h]
    \begin{center}
        \includegraphics[width=\textwidth]{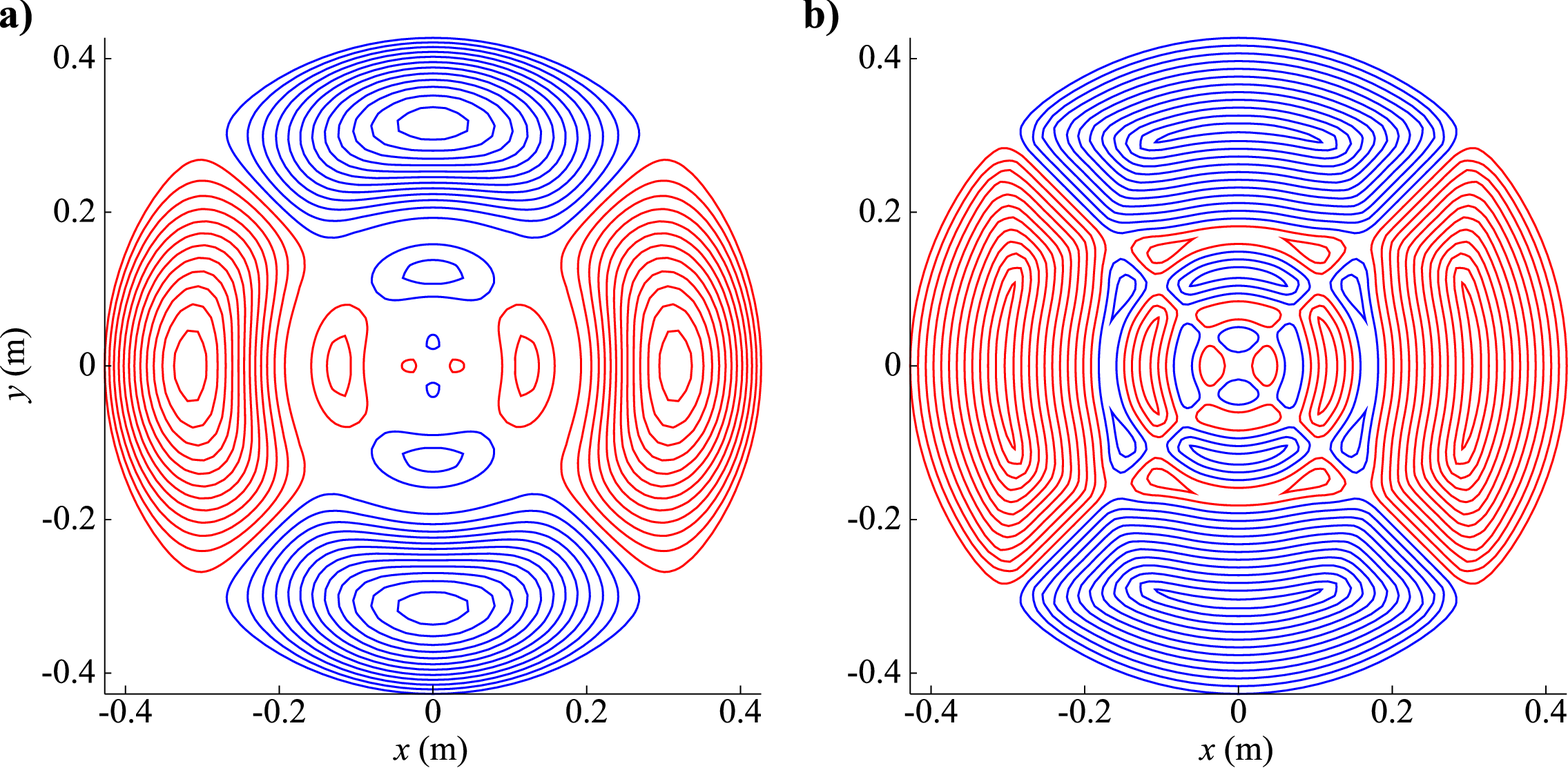}
    \end{center}
    \caption{ \label{fig:shim} One plane of the wire paths for the
    axisymmetric BEM designed biplanar X2-Y2 shim coils in the cases
    of \textbf{a)} min($P$) and \textbf{b)} \mmj. Red wires indicate
    reversed current flow with respect to blue.
    }
\end{figure}

The wire-paths of one plane of a \mmj\ X2-Y2 biplanar shim coil
are shown in Fig. \ref{fig:shim} \textbf{b)} next to an equivalent
min($P$) coil. Given a 4 mm wire spacing limit for construction,
the maximum achievable $\eta$ were 73.8 and 102.0
mTm$^{-2}$A$^{-1}$ for the min($P$) and \mmj\ coils using $N_c =
17$ and 20, respectively.

\section{Discussion}\label{sec:discussion}

This paper reports a method to directly minimize the maximum
current density for the magnetic field synthesis or coil design
problem. It focuses on the design of gradient and shim coils for
MRI applications, but can be considered as a general approach to
coil design; it may prove useful outside the realm of gradient and
shim coil design. Superconducting magnet design (e.g. see Ref.
\cite{Xu2000}) is one such application that merits some comment.
Although no experiments have yet been performed on magnet design
using this minimax optimization, it might be employed to design
magnets with reduced peak current density and/or peak magnetic
field in the conductors, for example. For practical designs,
$\ell^1$-norm minimisation of the current density may be
incorporated to yield low peak, yet sparse current density
designs.

Previous gradient coil design methods have very effectively
minimized the stored energy \cite{Turner1988} and power
dissipation \cite{Turner1993} subject to the production of
magnetic fields of a prescribed accuracy. Other approaches that
lower the maximum current density have been presented
\cite{Peeren2003a,Poole2008b,Zhu2008}, but none can be shown to be
optimal in terms of minimax current density. We used the adaptive
regularisation technique \cite{Poole2008b} to design shielded
X-gradient coils shown in Fig. \ref{fig:wbxgrad} and found that it
could achieve a minimum wire spacing of 8.7 mm. The \mmj\
algorithm achieved 9.3 mm wire spacing indicating that adaptive
regularisation works well, but cannot maximally spread the wires.
The reason for the difficulty in achieving truly minimax current
density coils is that such a term is non-differentiable with
respect to the solution variables. It is surely possible to insert
such a non-differentiable term into a stochastic optimization
technique such as a genetic algorithm \cite{Fisher1995} or
simulated annealing \cite{Crozier1993}, but it is expected that
such methods would require very long computing times and converge
to a solution that is not necessary the global one. Here, the
maximum operation is expressed as the infinity norm (also known as
the uniform or Chebychev norm), $\|\cdot\|_\infty$, smoothed and
converted to its norm-dual, the $\ell^1$-norm.

The time required for this algorithm to converge varies widely on
the size of the $(S \times N)$ current density matrices, $J_x$,
$J_y$ and $J_z$. For the Turner-Carlson approach \cite{Turner1986,
Turner1988, Stekly1985, Carlson1992} with $ N = 32$ and $S=441$
convergence was obtained in 23 seconds. However, for $N=1985$ and
$S=4096$ with the triangular BEM it took 480 minutes. It should be
noted that for $\delta = 0$, the solution is obtained in less than
a second as just one matrix inversion is needed
\cite{Lemdiasov2005, Poole2007}. This illustrates the need to take
into account any symmetry that might be present in the system to
lower the number of free variables and speed up the calculation.
For a defined maximum field error, the design process needs to be
repeated in order to obtain the ideal trade-off parameter,
$\delta$. It is hoped that the amount of user input and
computational burden can be reduced by describing the problem as a
constrained one in which maximum field error is a user-definable
constraint.

The length of an unshielded X gradient coil was varied such that
the length-to-diameter ratio, $l/d$, ranged from 0.92 to 2.63. The
performance of the coils designed with min($W$), min($P$) and
\mmj\ were evaluated and Figure \ref{fig:length} shows the
dependencies of $\eta^2/L$, $\eta^2/R$ and $\eta w$ on coil
length. $\eta^2/L$ characterises the power requirements of the
driving amplifier and $\eta^2/R$ characterises the total amount of
heat generated by the coil where higher values indicate better
performance in both cases. $\eta w$ on the other hand
characterises the maximum field strength that can be obtained
irrespective of inductance or resistance for a given minimum wire
spacing. Several interesting behaviours are apparent from studying
the data in Figure \ref{fig:length}. First, when designing a coil
with min($W$) it will have the highest $\eta^2/L$ value. Likewise,
a min($P$) coil will have the highest $\eta^2/R$ value and \mmj\
coils will have the highest $\eta w$. This is to be expected.
min($W$) and min($P$) designed coils have very similar $\eta^2/L$
values and slightly different $\eta^2/R$ values, with the \mmj\
coils possessing lower values of these two FOMs. In fact, for very
long coils ($l/d \gtrsim 2.2$), the value of $\eta^2/L$ and
$\eta^2/R$ actually decreases for the \mmj\ coils. However, the
\mmj\ coils possess a $\eta w$ value that is considerably larger
than that of the min($W$) and min($P$) designed coils. Figure
\ref{fig:length} c) shows that for min($W$) and min($P$) coils
with $l/d \gtrsim 2$ the value of $\eta w$ becomes flat. This
happens when the region of max$|j|$ occurs approximately at the
end of the ROI and not at the end of the coil. This indicates that
the length of the coil surface is no longer restricting the
maximum achievable field strength when $l/d \gtrsim 2$. $\eta w$
appears to be tending to a particular value for long \mmj\ coils
that is approximately 1.5 to 2 times larger than the other coils.
Unlike a previous approach \cite{Poole2008b}, minimax current
density coils are dramatically different from their Tikhonov
regularized counterparts even for long cylindrical coils.

It is known that a unique solution is found when ill-posed
problems are solved with Tikhonov regularisation, which is the
case for min($W$) and min($P$). It is not known if a unique
solution results from including the \mmj\ term in the functional.
It is suspected by the authors that there is no unique solution,
but more theoretical analysis is required to establish this.
Figure \ref{fig:gammaDelta} shows the behavior when $\beta = 0$
and both $\gamma$ and $\delta$ are finite, i.e. as min($P$) is
traded for \mmj\ in the optimisation. The $\delta$ value required
to maintain a constant field error is inversely related to
$\gamma$, as expected. $\gamma$ and $\delta$ values are similar
for $N=32$ and 200 sinusoidal basis functions. Figure
\ref{fig:gammaDelta} \textbf{b)} shows the variation of the power
FOM, $\eta^2/R$, as this trade-off happens. It is evident from
these data that by adding a small amount of $\gamma$ to the \mmj\
coil, a sharp increase in $\eta^2/R$ can be effected at the
expense of very little decrease in $\eta w$. When $N=32$ more
smoothness is enforced by the basis functions and $P$ is limited.
Hence $\eta^2/R$ is lower when more basis functions are used,
indicated by \ding{173} and \ding{174} on Figure
\ref{fig:gammaDelta} b), but $\eta w$ is also limited. This
difference is also evident in Figures \ref{fig:sosgrads} c) and
d). Conversely, Figure \ref{fig:gammaDelta} c) shows that by
adding a small amount of $\delta$ to the min($P$) coil, a large
increase in $\eta w$ can be achieved with only a small change in
the power dissipation of the coil. It is not surprising to observe
that both min($P$) coils with $N=32$ and 200 are essentially the
same since min($P$) coils favour low spatial frequencies in
$\psi(\mathbf{r})$. It is clearly possible to choose any solution
on the continuum from min($P$) to \mmj. Although not presented in
this paper, it is also possible to trade min($W$) with \mmj\ along
a similar continuum.

From Figure \ref{fig:sosgrads} it can be seen that the min($W$)
coil possesses an area with the highest current density at the
ends of the primary with the wires of the power minimized coil
being more spread, as expected \cite{Turner1993}.
$\psi(\mathbf{r})$ conforms to the usual $\cos{\phi}$ dependence
for the min($W$) and min($P$) coils despite no such constraint.
However, this leads to regions of higher current density at $\phi
= 0$ which is optimally dispersed in the \mmj\ coils. Deviation
from the $\cos{\phi}$ behavior at the ends of the coils means that
spherical harmonic fields of higher degree will be introduced in
the ROI. These high degrees are then cancelled by
$\psi(\mathbf{r})$ variations closer to the ROI.

Incorporating active magnetic shielding \cite{Mansfield1986a} into
the functional is a simple matter since it can be written in a
similar form as the target field term. Actively-shielded
X-gradient coils were designed using the same geometry as appears
in Reference \cite{Shvartsman2007}. Figure \ref{fig:wbxgrad} shows
one quadrant each of the primary and shield wire paths of the
min($W$), min($P$), \mmj\ and min($P$ \& max$|j|$). It is
interesting to note that the shield coil for the purely \mmj\ coil
has unnecessary current density with many reversed turns, Figure
\ref{fig:wbxgrad} f). This results from the fact that there is no
penalty for extra current density when $\beta$ and $\gamma$ are
zero. By incorporating a small amount of $\gamma$, this
impractical design is very effectively converted into a highly
practical design with smooth wire paths, low resistance and very
well spread wires, Figure \ref{fig:wbxgrad} h). It would also be
simple to have different $\gamma$ and $\delta$ values for primary
and shield coils. The wire paths in Figure \ref{fig:wbxgrad} e)
show the tendency of \mmj\ in an extreme case where right-angular
corners appear in the design.

Similar, but less pronounced effects were observed from the
results of the Z-gradient coil of identical geometry. Figure
\ref{fig:wbzgrad} shows the stream functions along the
$z$-direction for the coils designed with max$(\Delta B_z) = 5 \pm
0.01$ \%. The magnitude of the current density (in this case the
steepness of the slope of the stream-function) is almost the same
in all parts of the coil. This again leads to unnecessarily large
amounts of current density on the shield coil, which is easily
removed by the addition of a small value of $\gamma$. The combined
min($P$ \& max$|j|$) Z-gradient coil exhibits large wire spacing
and marginally increased resistance when compared to the min($P$)
coil. The problems associated with high current densities are less
severe when compared to those of X-gradient coils, but this
approach may be more useful for zonal shim coils of higher order.

In a final example, an axisymmetric BEM was used to design
biplanar X2-Y2 shim coils. Rotational symmetry of the system about
the $z$-axis is assumed. $\psi(\mathbf{r})$ is spectrally
decomposed in $\phi$ and spatially in $\rho$ and $z$.
Qualitatively, it can be seen from Figure \ref{fig:shim} that the
\mmj\ coil used all the space provided, whereas the min($P$) coil
forced wires to be very smooth. For a fixed max$(\Delta B_z) = 10
\pm 0.01$ \% and $w \geq 4$ mm, $\eta$ is 38\% higher for the
\mmj\ coil. The construction of such coils may be made slightly
more complex by the additional loops in the design. Again, a
combined min($P$ \& max$|j|$) coil might provide a good balance
between simplicity and efficiency.

\section{Conclusion}

It has been shown that the magnetostatic field synthesis problem
can be solved for coils with minimax current density. The problem
was solved in the present study with three different boundary
element methods to illustrate the generality of the approach. It
can therefore be used to synthesize any physically realistic
magnetic field with currents flowing on arbitrary surfaces.
Alternatively, the time needed to solve the problem can be
dramatically reduced by assuming some degree of symmetry. Coils
with minimax current density possessed increased resistance and
inductance for the same field error and in some cases had high
current densities in regions known to naturally require only low
current density. More practical coils were obtained with a mixture
of power and maximum current density minimization. Such coils are
characterized by low inductance and resistance, but also a large
spacing between all the wires of the coil. This spreading of wires
may be used to increase the efficiency of the coil by permitting
extra turns to be added, reduce turn-to-turn eddy current effects,
reduce localized heating in the coil, and design coils that are
easier to manufacture and/or are extremely short. Moreover, it
allows a coil designer to explore a new range of optimal solutions
to the field synthesis problem. The resultant coils must be judged
by figures-of-merit appropriate to the desired characteristics of
the coil, since field accuracy, gradient efficiency, stored
energy, power dissipation and maximum current density may all be
traded for each other.

\section{Acknowledgements}

This work was performed for MedTeQ, a Queensland smart state
funded research centre. The authors wish to thank Prof. Richard
Bowtell for his invaluable assistance with preliminary work on the
coil design methods used in this study.
%
%\newpage
\appendix
\section{Sum-of-Sinusoids}\label{sec:appA}

In this appendix we present for completeness the formulation for
calculating the matrices in Equations \eref{eq:jdisc} to
\eref{eq:matR} necessary for implementing the minimax current
density algorithm with sinusoidal stream-function basis functions.
In this case, the coil surface is assumed to be a finite-length,
$l$, cylinder of radius $a$ with its axis of symmetry oriented in
the $z$ direction. The stream-function of the current density,
$\psi(\mathbf{r})$, is spectrally decomposed and assumed to be a
finite weighted sum of truncated sinusoidal functions in $z$ and
$\phi$;

\begin{equation}\label{eq:sos0}
\psi(\phi,z) = \sum_{m^\prime=1}^{M^\prime}
\sum_{n^\prime=1}^{N^\prime} \lambda_{m^\prime n^\prime}
\hat{\psi}_{m^\prime n^\prime}(\phi,z)
\end{equation}

\noindent where

\begin{equation}\label{eq:sos1}
\hat{\psi}_{m^\prime n^\prime}(\phi,z) = \cases{
\sin{\left(\frac{2 \pi n^\prime}{l}z\right)} \cos{\left(
(2m^\prime - 1) \phi \right)}
& if $|z| \leq \frac{l}{2}$\\
0& if $|z| > \frac{l}{2}$ \\}.
\end{equation}

The prime indicates the difference between $n$ used in the main
algorithm and the order, $n^\prime$, and degree, $m^\prime$, of
the sinusoid. \Eref{eq:sfdisc} is obtained by a reordering of the
weights $\lambda_{m^\prime n^\prime}$ as $\psi_n$. It restricts
the magnetic field to be antisymmetric in the $x$-direction and
symmetric in the $z$-direction. Extra basis functions that are
symmetric in $x$ and antisymmetric in $z$ can be included in order
to remove the inherent field symmetry enforcement
\cite{Forbes2001, Forbes2002, Forbes2003}, but this is not
required in the present study since we are designing an X-gradient
coil and know that $\psi(\mathbf{r}) = 0$ at $|\phi| =
\frac{\pi}{2}$. Due to this enforced symmetry it is known that the
net torque experienced by the coil is zero.

The current density on the coil surface has $J_\phi$- and
$J_z$-components that are

\begin{equation}\label{eq:sos2}
J_\phi(\phi,z) = \frac{\partial \psi}{\partial z} =
\sum_{m^\prime}^{M^\prime} \sum_{n^\prime}^{N^\prime}
\lambda_{m^\prime n^\prime} \frac{2 \pi n^\prime}{l}
\cos{\left(\frac{2 \pi n^\prime}{l}z\right)} \cos{\left(
(2m^\prime - 1) \phi \right)}
\end{equation}

\begin{equation}\label{eq:sos2}
J_z(\phi,z) = \frac{-\partial \psi}{a\partial \phi} =
\sum_{m^\prime}^{M^\prime} \sum_{n^\prime}^{N^\prime}
\lambda_{m^\prime n^\prime} \frac{m^\prime}{a} \sin{\left(\frac{2
\pi n^\prime}{l}z\right)} \sin{\left( (2m^\prime - 1) \phi
\right)}
\end{equation}

\noindent and are equivalent to \eref{eq:jdisc} and \eref{eq:matG}.% The

\section{Triangular Boundary Elements}\label{sec:appB}

A surface can be meshed as a series of $I$ triangular elements
with $N$ nodes at the corners of the triangles
\cite{Pissanetzky1992}. In this case, $\psi(\mathbf{r})$ is
piecewise-linear in each triangle and the stream-function values
at the node positions, $\psi_n$, define the whole stream-function;

\begin{equation}\label{eq:tri1}
\psi(\mathbf{r}) = \sum_{n=1}^N \psi_n \sum_{i=1}^I
\hat{\psi}_{ni}(\mathbf{r})
\end{equation}

\noindent where

\begin{equation}\label{eq:triang}
\hat{\psi}_{ni}(\mathbf{r}) = 1 - \frac{(\mathbf{r} -
\mathbf{r}_n) \cdot \mathbf{d}_{ni}}{|\mathbf{d}_{ni}|}
\end{equation}

\noindent if $\mathbf{r}$ is a point in triangle $i$ and $n$ is a
node of that triangle. $\hat{\psi}_{ni}(\mathbf{r}) = 0$
otherwise. It is possible to use higher order shape functions over
the triangle so long as they form a divergence-free basis
\cite{Cobos2006}.

The current density on the surface is found from \eref{eq:tri1},
\eref{eq:triang} and \eref{eq:sfdef} yielding \eref{eq:jdisc} and

\begin{equation}\label{eq:GxConstruct}
\hat{\mathbf{j}}_n(\mathbf{r}) = \sum_i^I
\mathbf{v}_{ni}(\mathbf{r}) = \sum_i^I
\frac{\mathbf{e}_{ni}}{2A_i}
\end{equation}

\noindent if $\mathbf{r}$ is a point in triangle $i$ and $n$ is a
node of that triangle. $\mathbf{v}_{ni}(\mathbf{r}) = 0$
otherwise. $A_i$ is the area of the triangle $i$ and
$\mathbf{e}_{ni}$ is the vector that describes the edge of the
$i$th triangle opposite the $n$th node. This demonstrates that the
current density is uniform over each element of the mesh and so
that there needs to be one current density sample for each $I$
triangles in order to fully characterise it. Therefore $s$ becomes
$i$ and $S$ is equal to the number of triangles, $I$.

%To find the magnetic field at $\mathbf{r}_h$ the same discrete
%equation \eref{eq:sos6} with the basis functions
%
%\begin{equation}\label{eq:trib}
%\hat{b}_n(\mathbf{r}_h) = \frac{\mu_0}{4 \pi} \sum_i^I \int_{S_i}
%\frac{v_{nix}(y_h - y) - v_{niy}(x_h - x)}{|\mathbf{r}_h -
%\mathbf{r}|^3} dS.
%\end{equation}
%
%The discrete equations \eref{eq:sos8} and \eref{eq:sos10} for the
%energy terms have the basis functions $L_{nn^\ast}$ and
%$R_{nn^\ast}$;
%
%\begin{equation}\label{eq:tril}
%L_{nn^\ast} = \frac{\mu_0}{4\pi} \sum_i^I \sum_{i^\ast}^I
%(\mathbf{v}_{ni} \cdot \mathbf{v}_{n^\ast i^\ast}) \int_{S_i}
%\int_{S_i^\ast} \frac{dS^\ast dS}{|\mathbf{r}_{ni} -
%\mathbf{r}_{n^\ast i^\ast}|}
%\end{equation}
%
%\begin{equation}\label{eq:trir}
%R_{nn^\ast} = \frac{\rho_{Cu}}{t} \sum_i^I \sum_{i^\ast}^I
%(\mathbf{v}_{ni} \cdot \mathbf{v}_{n^\ast i^\ast}) A_{ni}
%\end{equation}
%
%This BEM is free from any symmetry constraints and so the net
%torque acting on the coil must be calculated. From \eref{eq:M} and
%\eref{eq:jdisc} the following equations are arrived at for the
%$x$- and $y$-components of the torque vector, $\btau$:
%
%\begin{equation}\label{eq:tritx}
%    \tau_x = B_0 \sum_n^N \psi_n \sum_i^I \int_{S_i} v_{nix} z dS,
%\end{equation}
%
%\begin{equation}\label{eq:trity}
%    \tau_y = B_0 \sum_n^N \psi_n \sum_i^I \int_{S_i} v_{niy} z dS.
%\end{equation}
%
%These can be written as inner products \eref{eq:matM}.

\section{Axi-symmetric Boundary Elements}\label{sec:appC}

The axisymmetric BEM is used for coil supports that can be
described by surfaces of revolution about the $z$-axis. Each
``node'', $n^\prime$, of this surface is in fact a circle in the
$xy$-plane and defined by its radius, $\rho_{n^\prime}$ and axial
position $z_{n^\prime}$. There may be a conical element, $i$,
either side of each node, labelled $+$ and $-$. A local
coordinate, $\zeta(\rho,z)$, is defined for each element that is 0
at one end and 1 at the other. Positions on these two conical
surfaces are

\begin{equation}\label{eq:posPlus}
\begin{array}{c}
  \mathbf{r}_n^+(\zeta, \phi) =  \left(\!\!\!%
\begin{array}{c}
  \rho_n\\
  \phi\\
  z_n\\
\end{array}\!\!\!%
\right) + \zeta \left(\!\!\!%
\begin{array}{c}
  \rho_{n+1} - \rho_n\\
  0\\
  z_{n+1} - z_n\\
\end{array}\!\!\!%
\right) \\
\\
  \mathbf{r}_n^-(\zeta, \phi) = \left(\!\!\!%
\begin{array}{c}
  \rho_{n-1}\\
  \phi\\
  z_{n-1}\\
\end{array}\!\!\!%
\right) + \zeta \left(\!\!\!%
\begin{array}{c}
  \rho_n - \rho_{n-1}\\
  0\\
  z_n - z_{n-1}\\
\end{array}\!\!\!%
\right) \\
\\
\end{array}
\end{equation}

The surface and therefore $\psi(\mathbf{r})$ are parameterised in
$\rho$ and $z$. $\psi(\mathbf{r})$ is decomposed spectrally
parameterised in the $\phi$-direction as a sum of sinusoids. In
section \S \ref{sec:methShim} an X2-Y2 shim coil is designed that
has a target magnetic field with 2-fold rotational symmetry about
$z$. Therefore, $\psi(\mathbf{r})$ is restricted in the
$\phi$-direction to take the form $\cos{(2(2m^\prime - 1)\phi)}$.
As described in \ref{sec:appA}, the basis function weights,
$\lambda_{m^\prime n^\prime}$ can be reordered to comply with the
vector arrangement, $\psi$, in the main algorithm.

\begin{equation}\label{eq:axsymSF}
\psi(\zeta,\phi) = \sum_{m^\prime}^{M^\prime}
\sum_{n^\prime}^{N^\prime} \lambda_{m^\prime n^\prime} \sum_i^I
\hat{\psi}_{m^\prime n^\prime i} (\zeta) \cos{(2(2m^\prime -
1)\phi)}
\end{equation}

\noindent where

\begin{equation}\label{eq:axsymSFBFp}
\hat{\psi}_{m^\prime n^\prime i} (\zeta) = (1-\zeta)
\end{equation}

\noindent if $i$ is on the positive side of $n^\prime$, for $0
\leq \zeta \leq 1$,

\begin{equation}\label{eq:axsymSFBFn}
\hat{\psi}_{m^\prime n^\prime i} (\zeta) = \zeta
\end{equation}

\noindent if $i$ is on the negative side of $n^\prime$, for $0
\leq \zeta \leq 1$ and $\hat{\psi}_{m^\prime n^\prime i} (\zeta) =
0$ otherwise.

The stream-function \eref{eq:sfdef} is applied to obtain, after
considerable amounts of algebra, the discretised current density

\begin{equation}\label{eq:axsymCD}
\mathbf{J}(\mathbf{r}) = \sum_{m^\prime}^{M^\prime}
\sum_{n^\prime}^{N^\prime} \lambda_{m^\prime n^\prime} \sum_i^I
\mathbf{v}_{m^\prime n^\prime i} (\mathbf{r})
\end{equation}

\noindent where

\begin{eqnarray}\label{eq:CDBFcart1}
%\begin{array}{lll}
\mathbf{v}_{m^\prime n^\prime i} (\mathbf{r}) &=
\left[\frac{\cos(2(2 m^\prime - 1)\phi)
\sin\phi}{\sqrt{(\rho_{n+1} - \rho_n)^2+(z_{n+1} - z_n)^2} }
\right. \nonumber \\
&\left.+ \frac{m^\prime (1-\zeta) (\rho_{n+1} - \rho_n) \sin(2(2
m^\prime -1)\phi)\cos \phi}{\sqrt{(\rho_{n+1} - \rho_n)^2+(z_{n+1}
- z_n)^2} (\rho_n + \zeta (\rho_{n+1} - \rho_n))}\right]
\mathbf{\hat{x}}\nonumber \\
&+ \left[\frac{-\cos(2(2 m^\prime - 1)\phi)
\cos\phi}{\sqrt{(\rho_{n+1} - \rho_n)^2+(z_{n+1} - z_n)^2} }
\right. \nonumber \\
&\left.+ \frac{m^\prime (1-\zeta) (\rho_{n+1} - \rho_n) \sin(2(2
m^\prime -1)\phi)\sin \phi}{\sqrt{(\rho_{n+1} - \rho_n)^2+(z_{n+1}
- z_n)^2} (\rho_n + \zeta (\rho_{n+1} - \rho_n))}\right]
\mathbf{\hat{y}}\nonumber \\
&+ \left[\frac{m^\prime(1-\zeta) (z_{n+1} - z_n) \sin(2(2 m^\prime
- 1)\phi)}{\sqrt{(\rho_{n+1} - \rho_n)^2+(z_{n+1} - z_n)^2}(\rho_n
+ \zeta (\rho_{n+1} - \rho_n))}\right]  \mathbf{\hat{z}}
%\end{array}
\end{eqnarray}

\noindent if $i$ is on the positive side of $n^\prime$, for $0
\leq \zeta \leq 1$ and $0 \leq \phi < 2 \pi$,

\begin{eqnarray}\label{eq:CDBFcart2}
%\begin{array}{lll}
\mathbf{v}_{m^\prime n^\prime i} (\mathbf{r}) &=
\left[\frac{-\cos(2(2 m^\prime - 1)\phi) \sin\phi}{\sqrt{(\rho_{n}
- \rho_{n-1})^2+(z_{n} - z_{n-1})^2} }
\right. \nonumber \\
&\left.+ \frac{m^\prime \zeta (\rho_{n} - \rho_{n-1}) \sin(2(2
m^\prime -1)\phi)\cos \phi}{\sqrt{(\rho_{n} - \rho_{n-1})^2+(z_{n}
- z_{n-1})^2} (\rho_{n-1} + \zeta (\rho_{n} - \rho_{n-1}))}\right]
\mathbf{\hat{x}}\nonumber \\
&+ \left[\frac{\cos(2(2 m^\prime - 1)\phi)
\cos\phi}{\sqrt{(\rho_{n} - \rho_{n-1})^2+(z_{n} - z_{n-1})^2} }
\right. \nonumber \\
&\left.+ \frac{m^\prime \zeta (\rho_{n} - \rho_{n-1}) \sin(2(2
m^\prime -1)\phi)\sin \phi}{\sqrt{(\rho_{n} - \rho_{n-1})^2+(z_{n}
- z_{n-1})^2} (\rho_{n-1} + \zeta (\rho_{n} - \rho_{n-1}))}\right]
\mathbf{\hat{y}}\nonumber \\
&+ \left[\frac{m^\prime \zeta (z_{n} - z_{n-1}) \sin(2(2 m^\prime
- 1)\phi)}{\sqrt{(\rho_{n} - \rho_{n-1})^2+(z_{n} -
z_{n-1})^2}(\rho_{n-1} + \zeta (\rho_{n} - \rho_{n-1}))}\right]
\mathbf{\hat{z}}
%\end{array}
\end{eqnarray}

\noindent if $i$ is on the negative side of $n^\prime$, for $0
\leq \zeta \leq 1$ and $0 \leq \phi < 2 \pi$ and
$\mathbf{v}_{m^\prime n^\prime i} (\mathbf{r})  = 0$ otherwise.

%The equations for the magnetic field, stored energy, power
%dissipation and torque in the axisymmetric BEM are extremely
%lengthy and in the interest of brevity, the reader is referred to
%the thesis of Peeren \cite{Peeren2003a}, where this is presented
%in detail.

%
\section*{References}
\bibliographystyle{unsrt}
\bibliography{./repbib}

\end{document}